\documentclass[prl,aps,twocolumn,floatfix,amsmath,amssymb,superscriptaddress]{revtex4-1}
\usepackage{latexsym}
\usepackage{graphicx}
\usepackage{tabularx}
\usepackage{times}
\usepackage[usenames,dvipsnames]{color}
\usepackage{amsmath}
\usepackage{dcolumn}
\usepackage{gensymb}
\usepackage{dsfont}
\usepackage{soul}
\usepackage{latexsym,amsmath,amssymb,bm,euscript}
\usepackage[
	colorlinks=true,
	urlcolor=BlueViolet,
 citecolor=Maroon,
	plainpages=false,
 	pdfpagelabels,
 	bookmarksnumbered
 	]{hyperref}
\usepackage{float}

\definecolor{gray}{rgb}{0.4,.4,0.4}
\definecolor{purple}{rgb}{0.6,.0,0.6}
\definecolor{darkgreen}{rgb}{0.0,.6,0.}

\def\beq{\begin{equation}}
\def\eeq{\end{equation}}

\begin{document}

\title{Tunable spin-orbit coupling and symmetry-protected edge states in graphene/WS$_2$}

\author{Bowen Yang}
\affiliation{Department of Physics and Astronomy, University of California, Riverside, CA 92521, USA}

\author{Min-Feng Tu}
\affiliation{Department of Physics and Institute for Quantum Information and Matter, California Institute of Technology, Pasadena, CA 91125, USA}

\author{Jeongwoo Kim}
\affiliation{Department of Physics and Astronomy, University of California, Irvine, CA 92697, USA}

\author{Yong Wu}
\affiliation{Department of Physics and Astronomy, University of California, Riverside, CA 92521, USA}

\author{Hui Wang}
\affiliation{Department of Physics and Astronomy, University of California, Irvine, CA 92697, USA}

\author{Jason Alicea}
\affiliation{Department of Physics and Institute for Quantum Information and Matter, California Institute of Technology, Pasadena, CA 91125, USA}
\affiliation{Walter Burke Institute for Theoretical Physics, California Institute of Technology, Pasadena, CA 91125, USA}

\author{Ruqian Wu}
\affiliation{Department of Physics and Astronomy, University of California, Irvine, CA 92697, USA}

\author{Marc Bockrath}
\affiliation{Department of Physics and Astronomy, University of California, Riverside, CA 92521, USA}

\author{Jing Shi}
\affiliation{Department of Physics and Astronomy, University of California, Riverside, CA 92521, USA}

\begin{abstract}

We demonstrate clear weak anti-localization (WAL) effect arising from induced Rashba spin-orbit coupling (SOC) in WS$_2$-covered single-layer and bilayer graphene devices. Contrary to the uncovered region of a shared single-layer graphene flake, WAL in WS$_2$-covered graphene occurs over a wide range of carrier densities on both electron and hole sides.  At high carrier densities, we estimate the Rashba SOC relaxation rate to be $\sim 0.2 \rm{ps^{-1}}$ and show that it can be tuned by transverse electric fields. In addition to the  Rashba SOC, we also predict the existence of a`valley-Zeeman' SOC from first-principles calculations. The interplay between these two SOC's can open a non-topological but interesting gap in graphene; in particular, zigzag boundaries host four sub-gap edge states protected by time-reversal and crystalline symmetries. The graphene/WS$_2$ system provides a possible platform for these novel edge states.

\end{abstract}

\maketitle

{\bf \emph{Introduction.}}~Electron pseudospin in graphene and the associated chirality yield remarkable transport consequences including the half-integer quantum Hall effect \cite{Yuanbo05} and intrinsic weak anti-localization (WAL) \cite{Ando98}. Physical spin, by contrast, is often largely a spectator that couples weakly to momentum due to carbon's low mass, leading to much longer spin diffusion lengths ( $>1\,\mu\rm{m}$ at room temperature) than normal conductors \cite{Tombros07,Han11}. Graphene's extremely weak spin-orbit coupling (SOC) clearly has merits, yet greatly hinders the observation of important spin-dependent phenomena including the quantum spin Hall effect \cite{Kane05} and quantum anomalous Hall effect \cite{Qiao10}.
Fortunately, the open two-dimensional honeycomb structure allows tailoring the SOC strength by coupling to foreign atoms or materials \cite{Castro09, Weeks11, Ding11, Hu12, Ma12,Jin13, Ferreira14}.  Several experiments have pursued approaches of graphene hydrogenation \cite{Kaverzin15, Balakrishnan13} or fluorination \cite{Hong11} as well as heavy-adatom decoration \cite{Jia15, Chandni15}; these methods tend to decrease the transport quality, and moreover the induced SOC appears either difficult to reproduce \cite{Kaverzin15, Balakrishnan13} or to detect \cite{Hong11,Jia15,Chandni15}.  A different approach has recently been employed by several groups: placing graphene on target substrates featuring heavy atoms. Proximity to the substrates not only provides desirable properties such as ferromagnetic ordering and large SOC, but also reduces adverse effects on the target materials \cite{Wang15, Zilong15, Avsar14, Zhe15}. 

Here we employ magneto-conductance (MC) measurements to demonstrate enhanced SOC in graphene proximity-coupled to multilayer WS$_2$.  We quantify the spin-relaxation rate caused by Rashba SOC by fitting to WAL data, and further show that the Rashba strength is tunable via transverse electric fields.  Guided by first-principles calculations, we also predict that WS$_2$-covered graphene additionally features a prominent `valley-Zeeman' SOC that mimics a Zeeman field with opposite signs for the two valleys.  The interplay between these two SOC terms opens a non-topological gap at the Dirac point that supports symmetry-protected sub-gap edge states along certain boundaries.  
Though the gap is too small to be detected in our experiments, theory suggests that graphene/WS$_2$ may provide a simple model system for studying such an unusual gapped phase.

\begin{figure}[b]

\includegraphics[width=\columnwidth]{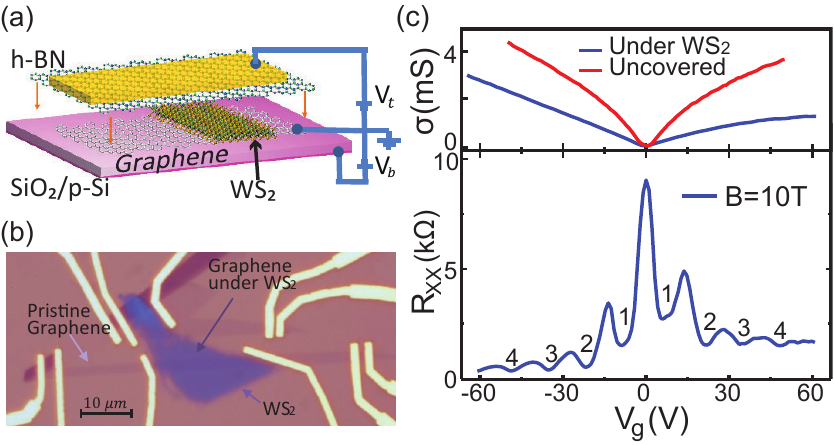}
\caption{(a) Device geometry. Bottom to top: SiO$_2$, graphene, WS$_2$, h-BN, and Au top-gate. h-BN serves as the dieletric for the top gate, and is transferred onto graphene/WS$_2$ after deposition of Au contacts. (b) Optical image of graphene/WS$_2$ before h-BN transfer. Two parallel graphene devices share the same WS$_2$ flake (dark blue) and each has WS$_2$-covered and uncovered
channels that can be probed independently. All single-layer-graphene data shown in this paper were taken from the lower device.  (c) Top: conductivity of uncovered (red) and WS$_2$-covered (blue) graphene devices. Bottom: Shubnikov-de Haas oscillations of WS$_2$-covered graphene measured at $2\, \rm{K}$ and $10\, \rm{T}$.  The evenly spaced peaks up to the $4$th order on both sides confirm the absence of carrier-density saturation.}
\label{fig:device}
\end{figure}

{\bf \emph{Experimental Setup.}}~ Figure~\ref{fig:device}(a) sketches the dual-gated graphene devices used in our study. Both single-layer graphene and multilayer WS$_2$ flakes were first exfoliated from their respective bulk materials and subsequently placed onto a Si/SiO$_2$ ($280\, \rm{nm}$) wafer. Since multilayer WS$_2$ flakes can be much thicker and are less likely deformed, we chose to transfer the WS$_2$ flake instead of graphene to avoid trapped bubbles in between, thereby yielding a larger effective overlap area. Figure~\ref{fig:device}(b) shows an optical image of the device prior to top-gate fabrication. Notice that only part of the graphene channel directly contacts with WS$_2$; the left uncovered channel serves as a control sample that allows direct comparison with the right part under WS$_2$ (dark blue).

Transport measurements were performed at 2K (unless specified otherwise) using a Quantum Design's Physical Property Measurement System.  Figure~\ref{fig:device}(c), top, shows the conductivity of graphene versus the back gate voltage. Interestingly, for both top and back gate sweeps, the device does not show the conductivity saturation (up to $\pm60\rm{V}$ with back gate) reported recently by other groups \cite{Avsar14,Zhe15}. Conductivity saturation in the latter studies was attributed to saturation in carrier density from either the large density of states associated with sulfide defects \cite{Avsar14} or screening by electrons in the WS$_2$/SiO$_2$ interface \cite{Zhe15}. In our WS$_2$-covered device, the lack of conductivity saturation on either side suggests that the Fermi level resides within the band gap of WS$_2$, consistent with our DFT calculations (see below).  The absence of the carrier density saturation in graphene is verified by Shubnikov-de Haas oscillations of the WS$_2$-covered graphene as a function of the gate voltage in a $10 \,\rm{T}$ magnetic field; see Figure~\ref{fig:device}(c). On both sides, the Landau Levels are evenly spaced up to the 4th level, indicating that the carrier density is proportional to the gate voltage. This property allows us to access the high-density regions, which is important for understanding the origin of enhanced SOC and accurately determining its strength. The field effect mobility, calculated from capacitance of the SiO$_2$ \cite{Chen08} layer, is higher in the uncovered graphene ($\sim 7000\, \rm{cm}^2\rm{s}^{-2}\rm{V}^{-1}$) than the WS$_2$-covered graphene (  $\sim 4000\, \rm{cm}^2\rm{s}^{-2}\rm{V}^{-1}$ on the hole side, and $\sim 2000\, \rm{cm}^2\rm{s}^{-2}\rm{V}^{-1}$ on the electron side). Despite the relatively low mobility, our devices manifest clear low-field magneto-conductance (MC) over a much larger carrier-density range than in previous studies \cite{Zhe15, Avsar14}.

\begin{figure}[ht]
\includegraphics[width=\columnwidth]{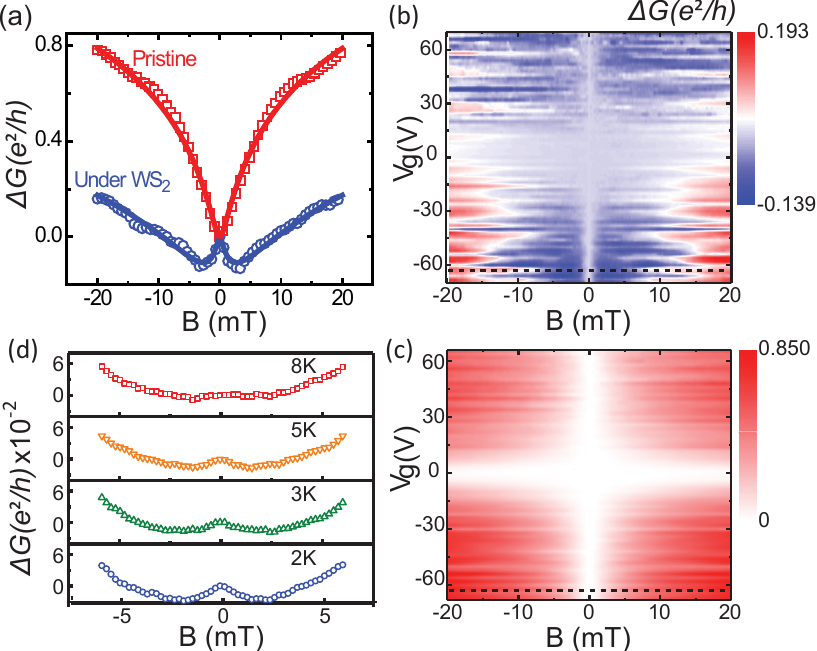}
\caption{(a) MC comparison between WS$_2$-covered (blue circles) and uncovered (red squares) graphene channels at carrier density $n= 5\times10^{12}\rm{cm}^{-2}$ [dotted lines in (b) and (c)]. Solid blue and red curves represent fits using Eq.~\eqref{eq:MC_SOC} and Ref.~\onlinecite{McCann06}, respectively. (b,c) Gate-voltage dependence of MC for (b) WS$_2$-covered and (c) uncovered devices.  The narrow white vertical region near $B = 0$ in (b) represents the WAL \emph{peak} in WS$_2$-covered graphene, whereas a WL \emph{dip} near $B = 0$ appears for all gate voltages in uncovered graphene (c). (d) Temperature dependence of the WAL in a bilayer graphene device, with carrier density $n= 8\times10^{12}\rm{cm}^{-2}$. }
\label{fig:MC}
\end{figure}

{\bf \emph{Rashba SOC Signature.}}~  
Due to its unusual chirality, graphene with smooth disorder is predicted to exhibit WAL \cite{Ando98}.  However, strong inter-valley scattering, which typically arises in ordinary-quality samples, suppresses the chirality-related WAL and generates weak localization (WL) \cite{McCann06, Tikhonenko08}.  Introducing strong Rashba SOC allows the spin relaxation rate $\tau_{\rm{R}}^{-1}$ to exceed the inelastic dephasing rate $\tau_{\phi}^{-1}$.  In this case, before quantum dephasing occurs the electron spin precesses around the effective magnetic field and acquires an additional $\pi$ phase in the interference \cite{McCann12}---reviving WAL due to spin. Intrinsic (Kane-Mele) and valley-Zeeman SOC terms, by contrast, break an effective time reversal symmetry and thus place the system in the unitary class (suppressed WL) \cite{McCann12}.

Figure~\ref{fig:MC}(a) contrasts the low-temperature MC $\Delta G = G-G(B=0)$ for uncovered and WS$_2$-covered devices at approximately the same carrier density, $n=-5\times 10^{12} \rm{cm}^{-2}$ [corresponding to the black dashed lines in Figs.~\ref{fig:MC}(b) and (c)].  The uncovered graphene shows WL as expected given the modest mobility. More interestingly, in WS$_2$-covered graphene the MC clearly exhibits the hallmark WAL feature at low fields.  In both cases this behavior persists over a broad range of gate voltages as shown in Figs.~\ref{fig:MC}(b) and (c).  The robust WAL feature appearing only in the WS$_2$-covered graphene---despite its lower mobility which naively further promotes WL---provides strong evidence of Rashba SOC acquired from WS$_2$ on both electron and hole sides.  This result differs qualitatively from the strongly asymmetric characteristic reported in Ref.~\onlinecite{Avsar14}; there the induced SOC was only observed on the electron side, which was attributed to the asymmetric density-of-states due to sulfur vacancies. 

To further confirm the proximity-induced SOC, we fabricated a WS$_2$-covered bilayer-graphene device.  Unlike in single-layer graphene, WL is expected independent of  inter-valley scattering strength in bilayer graphene due to its associated $2\pi$ Berry phase \cite{Gorbachev07}.  Consequently, the emergence of WAL in a bilayer graphene---which we indeed detect---gives direct evidence of Rashba SOC inherited from WS$_2$ (i.e., the competing pseudospin interpretation disappears here).  Figure~\ref{fig:MC}(d) shows the observed WAL feature in a bilayer-graphene device at different temperatures. Note that we only measure a clear WAL signature when the carrier density exceeds $\sim8\times 10^{12}\rm{cm}^{-2}$, suggesting that the dominant dephasing mechanism in bilayer graphene is electron-electron interaction.  In this scenario, increasing the carrier density suppresses dephasing, and WAL appears once the dephasing rate drops below the spin relaxation rate.  The WAL feature also disappears on raising temperature, due naturally to thermally enhanced dephasing.  

It is worth mentioning that the MC data shown in Figure~\ref{fig:MC} are from single field-sweep measurements, as opposed to an ensemble average \cite{Tikhonenko09, Zhe15} over many curves taken over a gate-voltage range corresponding to the Thouless energy. Our device length ($\sim 20 \mu\rm{m}\times 2 \mu \rm{ m}$) greatly exceeds the coherence length ($\sim 1 \mu \rm{ m}$); hence the conductivity self-averages resulting in suppressed universal conductance fluctuations (UCF) \cite{Datta}.

{\bf \emph{Quantitative Analysis.}}~When inter- and intra-valley scattering rates are much larger than the dephasing and spin relaxation rates, MC in graphene is well-described at low magnetic fields by the following expression from diagrammatic perturbation theory \cite{McCann12}:
\begin{eqnarray}
  \Delta G &=& \frac{-e^2}{2\pi h}\bigg{[}F\left(\frac{B}{B_{\phi}}\right) - F\left(\frac{B}{B_{\phi}+2B_{\rm{asy}}}\right)
  \nonumber \\
  &-&2F\left(\frac{B}{B_{\phi}+B_{\rm{asy}}+B_{\rm{sym}}}\right)\bigg{]}.
\label{eq:MC_SOC}
\end{eqnarray}
Here $F\left(z\right) = \ln\left(z\right)+\Psi\left(\frac{1}{2}+\frac{1}{z} \right)$ ($\Psi$ is the digamma function) and $B_{\phi,\rm{asy}, \rm{sym}}=\frac{\hbar}{4De} \tau^{-1}_{\phi,\rm{asy}, \rm{sym}}$ with $D$ the diffusion constant.  The spin relaxation rate $\tau_{\rm{asy}}^{-1}$ is determined by the $z\rightarrow -z$ asymmetric Rashba SOC $\lambda_{\rm{R}}$, i.e., $\tau_{\rm{asy}}^{-1}=\tau_{\rm{R}}^{-1}$, while $\tau_{\rm{sym}}^{-1}$ follows from those $z\rightarrow -z$ symmetric SOCs including the intrinsic SOC $\lambda_{\rm{I}}$, and valley-Zeeman SOC $\lambda_{\rm{VZ}}$.  (Additional SOC terms that may be present due to the system's low symmetry are assumed negligible for simplicity.)

The intrinsic SOC relaxation rate $\tau_{\rm{I}}^{-1}$ obeys the Elliot-Yafet mechanism \cite{Elliott54, Yafet63,Ochoa12}:  $\tau_{\rm{I}}^{-1}= \tau_e^{-1} \left(\lambda_{\rm{I}}^2/E_{\rm{F}}^2 \right)$, where $\tau_e^{-1}$ is the momentum relaxation rate and $E_{\rm{F}}$ is the Fermi energy.  This rate is thus negligibly small compared to the typical dephasing rate in graphene when $\lambda_{\rm{I}}^2/E_{\rm{F}}^2\ll1$. Here we deliberately focus on the high-carrier-density region ($n>4\times 10^{12}\rm{cm}^{-2}$ and $E_F>0.2 \rm{eV}$) where we can reasonably approximate $\tau^{-1}_{\rm{sym}} \approx 0$. 
The $\lambda_{\rm VZ}$ coupling meanwhile is inherited from WS$_2$ due to sublattice symmetry breaking \cite{Xiao12}.  Since this term imposes an opposite Zeeman field for the two valleys, it generates \emph{non-degenerate}, spin-polarized momentum eigenstates whose spin orientations do not relax (except due to the interplay with other SOCs).  Thus the valley-Zeeman SOC relaxation rate is also negligible.  With these assumptions only $\tau_\phi^{-1}$ and $\tau_{\rm{R}}^{-1}$ remain in Eq.~\eqref{eq:MC_SOC}, and both can be extracted by fitting to the experimental data [see, e.g., blue curve in Fig.~\ref{fig:MC}(a)].  

\begin{figure}[ht]
\includegraphics[width=\columnwidth]{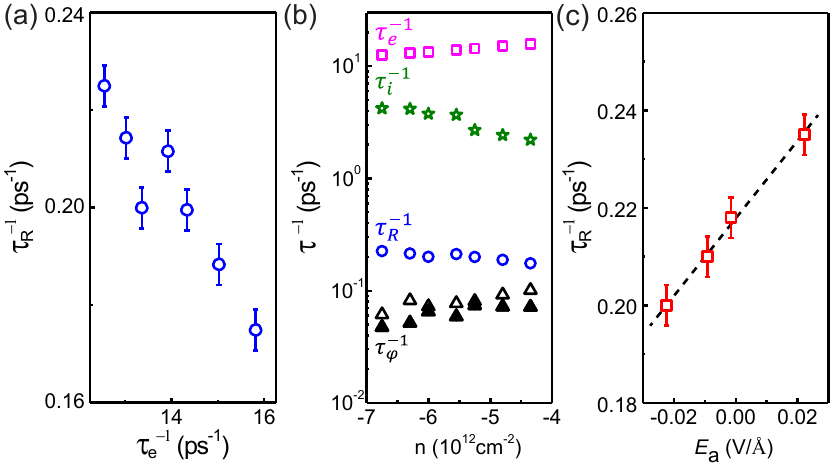}
\caption{(a) Rashba SOC relaxation rates as a function of the momentum scattering rates at carrier density $n=6.8\times 10^{12}\rm{cm^{-1}}$. Error bar indicate the fitting uncertainty. (b) Characteristic rates in WS$_{2}$-covered graphene as a function of the carrier density, except inter-valley scattering rates (stars) which are extracted from uncovered graphene.  Squares denote the momentum scattering rates, circles are the Rashba spin relaxation rates, and open (filled) triangles are the inelastic dephasing rates extracted from WAL (UCF). (c) Rashba SOC spin relaxation rates extracted at different transverse electric fields. Dashed line is a guide to the eyes. }
\label{fig:fitting}
\end{figure}

Figure~\ref{fig:fitting}(a) shows the resulting $\tau_{\rm{R}}^{-1}$ for WS$_2$-covered graphene as a function of the momentum scattering rate $\tau^{-1}_e$ calculated from the device mobility \cite{Rengel13}. As $\tau^{-1}_e$ increases, the Rashba SOC relaxation rate decreases almost monotonically, indicating that the spin relaxation is dominated by the Dyakonov-Perel mechanism \cite{Dyakonov72} [$\tau^{-1}_{\rm{R}}=2\tau_e(\lambda_{\rm{R}}/\hbar)^2$]. This behavior stands in marked contrast to standalone graphene, in which the Elliot-Yafet mechanism dominates spin relaxation over a broad range of carrier density \cite{Han11, Kaverzin15}. Furthermore, the spin relaxation rate of WS$_2$-covered graphene ($\tau_{\rm{R}}^{-1} \approx 0.2  \rm{ps}^{-1}$) exceeds that for standalone graphene (e.g., $\sim 3\times 10^{-3} \rm{ps}^{-1}$ \cite{Han11}) by two orders of magnitude---indicating strong SOC introduced by the proximity coupling with WS$_2$.

Figure~\ref{fig:fitting}(b) displays the density dependence of the characteristic relaxation rates.  All data correspond to WS$_2$-covered graphene except the inter-valley scattering rate $\tau_i^{-1}$.  The latter is extracted by fitting our WL data for uncovered graphene with the theory of Ref.~\onlinecite{McCann06} instead of Eq.~\eqref{eq:MC_SOC}; as an example, see the red curve in Fig.~\ref{fig:MC}(a).  [Equation~\eqref{eq:MC_SOC} can also provide a good fit for our low-field WL measurements in the absence of any SOC terms, but does not reveal $\tau_i^{-1}$.]  We assume that $\tau_i^{-1}$ inferred from uncovered graphene sets a lower bound for the corresponding rate in WS$_2$-covered graphene,which is quite natural given its lower mobility.  From Fig.~\ref{fig:fitting}(b) we then see that $\tau_i^{-1} \gg \tau_{\rm{R}}^{-1}$---a prerequisite for Eq.~\eqref{eq:MC_SOC}---is indeed satisfied for WS$_2$/graphene.  Moreover, the dephasing rate $\tau_\phi^{-1}$ can be extracted independently from the WAL, or the UCF by the autocorrelation function \cite{Lee87} (see Supplementary Material for details), and both methods agree quite well. These facts support the applicability of Eq.~\eqref{eq:MC_SOC} and suggest that the spin relaxation rates we extracted from the high-carrier density region are reliable.
 
Our dual-gated graphene device [Fig.~\ref{fig:device}(a)] allows us to study the influence of an applied transverse electric field on the Rashba SOC. In particular, the dual gate enables independent control of the carrier density (thus the momentum scattering rate) and the transverse electric field \cite{Yuanbo15}.  Figure~\ref{fig:fitting}(c) shows the spin relaxation rate $\tau^{-1}_{\rm{R}}$ extracted at fixed $\tau_e^{-1} = 12\rm{ps^{-1}}$ but at different transverse electric fields $E_a$ (for $E_a>0$ the field points from WS$_2$ to graphene).  Interestingly, $\tau^{-1}_{\rm{R}}$ increases monotonically with the applied field, changing by 18\% over the range -60V/300nm to 60V/300nm.  This increase can be interpreted as an enhancement of the Rashba SOC: The positive electric field lifts the graphene Dirac bands towards the WS$_2$ conduction bands \cite{Young09}; hence graphene's $\pi$ orbitals acquire a stronger hybridization with the tungsten $d$ orbitals, substantially strengthening Rashba SOC. 

\begin{figure}
\includegraphics[width=\columnwidth]{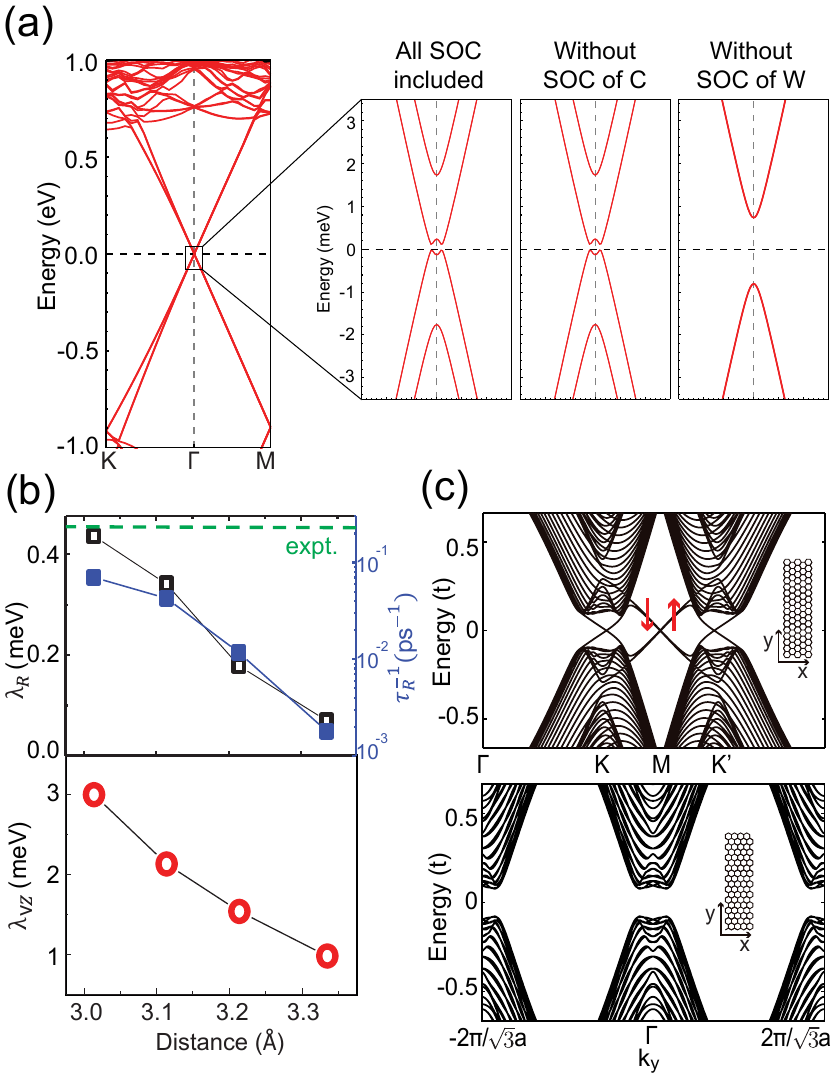}
\caption{(a) Calculated band structure for graphene/WS$_2$ heterostructure (left panel), and zoom-ins near the Dirac point (right panel) with SOC selectively included for different atoms. The leftmost zoom-in includes SOC for all atoms, while the middle and right exclude SOC for carbon and tungsten, respectively.  (b) Upper panel: calculated Rashba SOC and its associated spin relaxation rate versus interlayer distance. Green dashed line indicates the value of the experimentally extracted spin relaxation rate. Lower panel: interlayer distance dependence of valley-Zeeman SOC. (c) Energy bands for a graphene strip with zigzag edges (top) and armchair edges (bottom) using $\lambda_{\rm{VZ}}/t=0.3$ and $\lambda_{\rm{R}}/t=0.1$ ($t$ is the nearest-neighbor hopping strength for carbon).}
\label{fig:theory}
\end{figure}

{\bf \emph{Origin and implications of SOC.}}~To explain these experimental findings we performed density-functional theory (DFT) calculations using a large supercell in the lateral plane ($9\times9$ graphene on $7\times7$ WS$_2$) that minimizes the lattice mismatch ($0.35 \%$) between these two materials. With the van der Waals correction, the optimized interlayer distance is $3.34 \rm{\AA}$, and a small buckling ($< 0.08 \rm{\AA}$) is found in the graphene layer. The Dirac cones in Fig.~\ref{fig:theory}(a) still center around the Fermi level, indicating negligible charge transfer between WS$_2$ and graphene as seen experimentally [in all our devices the graphene is slightly p-doped ($n=0 \sim 1.5\times10^{12}\rm{{cm}^{-2}}$), as generally observed for $\rm{SiO_2}$ substrates].  The zoom-in of the band structure reveals a sizable spin splitting and a gap at the Dirac point due to SOC and the loss of sublattice symmetry. To diagnose the origin of the SOC terms, we adjust the SOC strength of each element selectively; see rightmost panels of Fig.~\ref{fig:theory}(a).  When SOC of carbon is excluded, the band structure remains essentially unchanged. However, eliminating the SOC for tungsten removes the spin splitting and yields a trivial mass gap, unrelated to SOC, that simply reflects the staggered sublattice potential induced by WS$_2$.   Enhanced SOC of graphene is thus primarily induced by hybridization with tungsten atoms.  

We analytically model our DFT results with the low-energy Hamiltonian
\begin{eqnarray}
H_{\rm eff}&=&\hbar v_{\rm{F}} (\tau_z \sigma_x p_x - \sigma_y p_y)+M\sigma_z \nonumber\\
&+& \lambda_{\rm{I}} \tau_z \sigma_z s_z +\lambda_{\rm{R}}(\tau_z \sigma_x s_y-\sigma_y s_x)+\lambda_{\rm{VZ}} \tau_z s_z.
\label{eq:hamiltonian}
\end{eqnarray}
The first line represents the standard Dirac theory supplemented by a staggered sublattice potential $M$, while the second encodes symmetry-allowed SOC terms \footnote{Note that inter-valley terms are excluded here even though the system is a $3n\times 3n$ superlattice; the supplementary material provides evidence that they are unimportant in this case.}.  
DFT bands near the Dirac point for the optimized structure can be well-fit using Eq.~\eqref{eq:hamiltonian} with the following parameters: $M = 0.79\rm{meV}$, $\lambda_{\rm{R}} = 0.03\rm{meV}$, $\lambda_{\rm{VZ}}= 0.96\rm{meV}$ and $ \lambda_{\rm{I}}\approx0\rm{meV}$.  

The fitted SOC strengths do, however, depend sensitively on the interlayer distance in the DFT simulations.  Figure~\ref{fig:theory}(b) presents the interlayer-distance dependence of the two dominant SOCs, $\lambda_{\rm{R}}$ and $\lambda_{\rm VZ}$.  The Rashba spin relaxation rates shown are calculated through $\tau^{-1}_{\rm{R}}=2\tau_e \left(\lambda_{\rm{R}}/\hbar\right)^2$, with a value $\tau_e = 12$ps$^{-1}$ comparable to that extracted from experiment. We find that DFT for the optimized structure underestimates the Rashba coupling $\lambda_{\rm{R}}$ seen experimentally, but that this difference can be mitigated by using $\sim5\%$ smaller interlayer distances.  This `correction' is not unreasonable given imperfections in our samples and the neglect of the weak force between graphene and WS$_2$ in DFT calculations.  The reduced distance also increases $\lambda_{\rm{VZ}}$ in DFT; its effect, however, is likely artificially enhanced by the use of a parallelogram supercell that breaks sublattice symmetry, which is arguably restored in an average sense by the incommensuration of real samples.  On the contrary, we expect that incommensuration more weakly impacts $\lambda_{\rm{R}}$, which only requires $z\rightarrow -z$ asymmetry. 

Together, these two SOCs open a gap at the neutrality point---$\lambda_{\rm{VZ}}$ lifts spin degeneracy while $\lambda_{\rm R}$ gaps the remaining carriers via spin-flip processes.  This gapped state is not a topological insulator (contrary to the reports of previous DFT studies \cite{Gmitra16, Zhe15}), as can be verified by the existence of an even number of counter-propagating edge states and explicit calculations of the topological invariant in a lattice model. 
Figure~\ref{fig:theory}(c) shows the tight-binding band structure for a strip with zigzag (top) and armchair (bottom) edges, including both $\lambda_{\rm R}$ and $\lambda_{\rm VZ}$ SOCs.  In the zigzag case two copies of edge states appear at $K$, $K'$ points due to band inversion, as observed in Ref.~\onlinecite{Gmitra16}, but two more edge states also appear at the $M$-point.  These edge states are protected by time reversal and crystalline symmetries, but do not have a topological origin. For an armchair geometry, no edge states appear. 

This gapped phase, while topologically trivial, exhibits edge-state properties that differ markedly from those of the valley Hall effect driven by an ordinary mass gap \cite{Xiao07, Yuanbo15}.  Both exhibit edge states along zigzag boundaries, but with very different spin polarizations.  For the SOC gap, the $\rm{M}$-point edge states exhibit out-of-plane spin polarization while those at $K$ and $K'$ exhibit in-plane polarization. 
In contrast, valley-Hall-effect edge modes are spin degenerate and thus do not naturally support spin currents.  The nontrivial spin structure for the edge modes in our problem, combined with the prospect of electrically tuning Rashba coupling and hence the band gap, underlie tantalizing applications for spintronics that warrant further pursuit. 

{\bf \emph{Conclusion.}}~We have demonstrated a dramatic and tunable enhancement of Rashba SOC in graphene by coupling to WS$_2$. In the high carrier-density region, we determined the Rashba coupling strength by analyzing the low-field MC.  First-principles calculations indicate that the induced SOC originates from the band hybridization between graphene $\pi$ orbitals and tungsten states.  The combination of Rashba and a theoretically predicted valley-Zeeman SOC creates novel edge states that are interesting to pursue further by engineering heterostructures with different substrates as well as improving the device mobilities.

{\bf \emph{Acknowledgments.}} We gratefully acknowledge Roger Mong for valuable discussions. 
This work was supported by the DOE BES award No.~DE-FG02-07ER46351 (BY and JS) and award No.~DE-FG02-05ER46237 (JW and RW); NSF through grant DMR-1341822 (MT and JA); the Caltech Institute for Quantum Information and Matter, an NSF Physics Frontiers Center with support of the Gordon and Betty Moore Foundation; and the Walter Burke Institute for Theoretical Physics at Caltech.  DFT simulations were performed on the U.S. Department of Energy Supercomputer Facility (NERSC). 

\bibliography{graphene,graphene_supp}

%%%%%%%%%%%%%%%%%%%%%%%%%%%%%%%%%%%%%%%%%%%%%%%%%%%%%%%%%%%%%%%%%%%%%%%%
\pagebreak
\onecolumngrid
\setcounter{equation}{0}
\setcounter{figure}{0}
\setcounter{page}{1}
\makeatletter
\renewcommand{\theequation}{S \arabic{equation}}
\renewcommand{\thefigure}{S \arabic{figure}}
%\renewcommand*{\bibnumfmt}[1]{[s#1]}
%\renewcommand*{\citenumfont}[1]{s#1}

%%%%%%%%%%%%%%%%%%%%%%%%%%%%%%%%%%%%%%%%%%%%%%%%%%%%%%%%%%%%%%%%%%%%%%%%
\section{\large Supplementary Information }

\section{Method and experimental setup}
The target WS$_2$ flake was identified with an optical microscope and then transferred to cover part of the long graphene channel. To promote adhesion, the wafer was annealed in O$_2$ at $300\celsius$ for 3 hours. Standard electron-beam lithography and electron-beam evaporation were used to connect the graphene with multiple $80 \rm{nm}$ thick Au electrodes. The electrodes allow independent four-terminal resistivity measurements in the covered and uncovered areas. After fabrication of electrodes, an h-BN flake was transferred to cover the whole device to serve as a top gate dielectric, followed by the top Au gate metal fabrication with similar electron-beam lithography procedures. No additional annealing was performed thereafter.

%%%%%%%%%%%%%%%%%%%%%%%%%%%%%%%%%%%%%%%%%%%%%%%%%%%%%%%%%%%%%%%%%%%%%%%%
\section{Reproducible WAL in a single-layer graphene on WS$_2$ device}
%\label{app.gauge}

\begin{figure}[H]
\includegraphics[width=\textwidth]{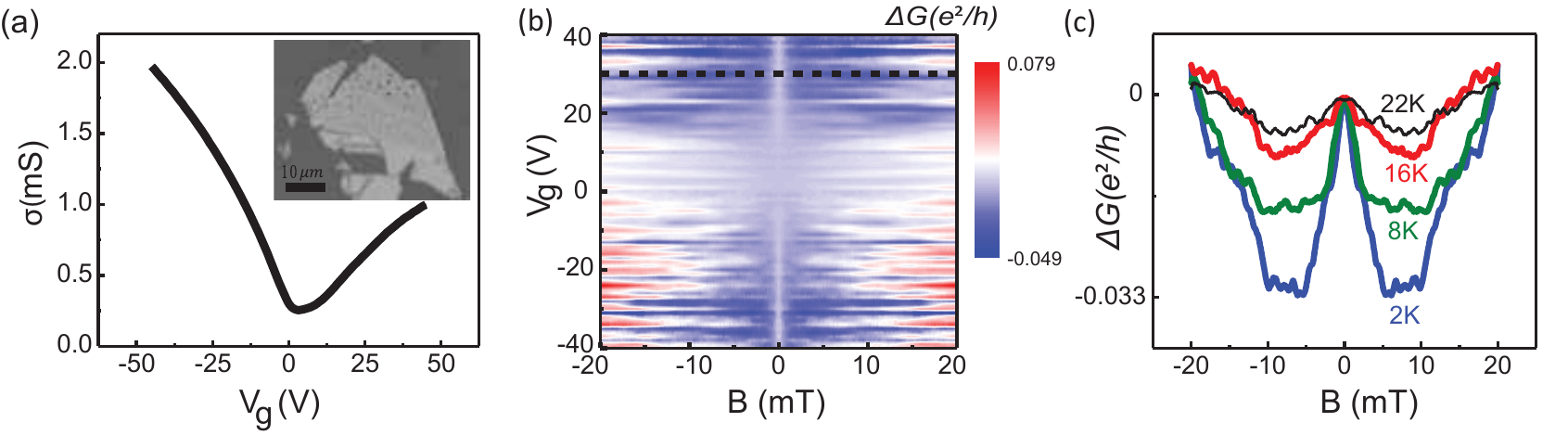}
\caption{(a) Conductivity versus gate voltage. Inset: SEM image of graphene on WS$_2$ before electrode fabrication. Bubbles are visible after graphene is transferred onto WS$_2$. (b) MC and its dependence on the temperature (c) at the trace corresponding to the black dotted line in (b). 
}
\label{fig:s1}
\end{figure}

Here we present a second device that has the similar characteristics as the first one shown in the main text. Instead of transferring WS$_2$ onto graphene, in this second sample we transferred graphene onto WS$_2$, in order to show that the absence of density saturation is independent of the transfer sequence and robustness of WAL exists. Figure~\ref{fig:s1}(a) shows the conductivity of graphene versus gate voltage. This device exhibits almost the same properties as the device in the main text, despite its lower mobility ($\sim 3000\rm{cm}^{-1}\rm{s}^{-1}\rm{V}^{-1}$ on the hole side, and $\sim 2000\rm{cm}^{-1}\rm{s}^{-1}\rm{V}^{-1}$    on the electron side), which is mainly due to the bubbles in the device. As clearly seen in the inset, bubbles (small black dots) appear across the graphene flake. We intentionally did not choose a bubble-free area in order to minimize the UCFs. Figure~\ref{fig:s1}(b) shows the WAL features observed in this device at different gate voltages. Compared to the first sample, MC is smallerhere.  We symmetrize the data to show clearer temperature dependence, as is shown in Figure~\ref{fig:s1}(c). Just as in the first device, WAL is present on both hole and electron sides and disappears quickly as temperature increases.

%%%%%%%%%%%%%%%%%%%%%%%%%%%%%%%%%%%%%%%%%%%%%%%%%%%%%%%%%%%%%%%%%%%%%%%%

\section{Temperature dependence of MC}

\begin{figure}[H]
\centering
\includegraphics[width=0.7\linewidth]{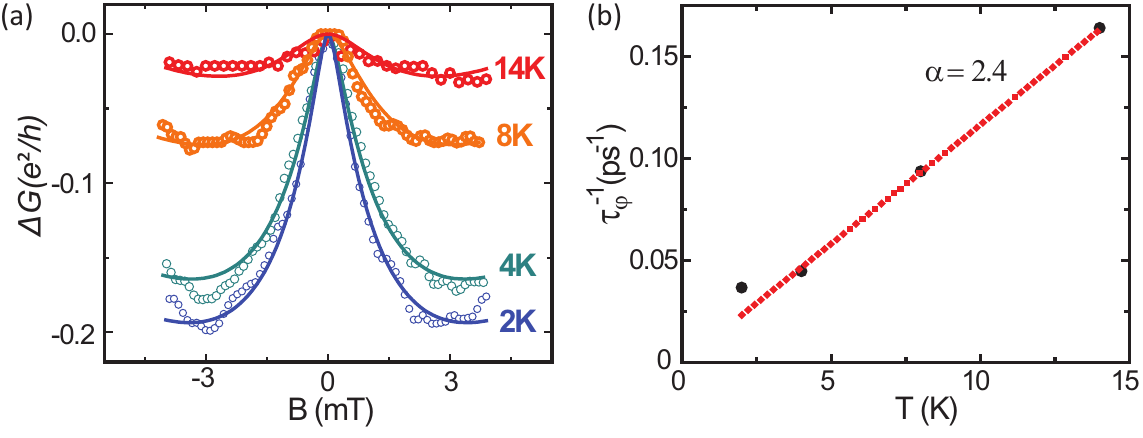}
\caption{(a) Temperature dependence of MC for WS$_2$-covered graphene at carrier density $n = 5\times10^{12}\rm{cm}^{-2}$. Solid lines are fits assuming a temperature-independent SOC rate. The dephasing rate extracted from the fitting is shown in (b). The temperature dependence of the dephasing rate (black dots) is approximately linear. The red dotted line is the dephasing rate calculated from Eq. \eqref{eq.phase_rate} with $\alpha = 2.4$. }
\label{fig:s3}
\end{figure}

Figure~\ref{fig:s3}(a) shows the temperature dependence of the MC data of the WS$_2$-covered graphene. Compared with the spin diffusion length and the SOC scattering rate \cite{Han11}, the inelastic dephasing rate $\tau_{\phi}^{-1}$ is much more sensitive to temperature at low temperatures; thus the dramatic decrease of the WAL signal can be primarily attributed to the significantly increased inelastic dephasing rate.  The extracted dephasing rate $\tau_{\phi}^{-1}$ as a function of temperature appears in Fig.~\ref{fig:s3}(b). We find that $\tau_{\phi}^{-1}$ obeys approximately a linear temperature dependence, which can be explained by the electron-electron scattering in the diffusive regime \cite{Altshuler82},
\begin{equation}
\tau_{\phi}^{-1}= \alpha \frac{k_B T}{2 E_F\tau_0}\ln\left(\frac{E_F\tau_0}{\hbar}\right),
\label{eq.phase_rate}
\end{equation}
where $\alpha$ is a correction coefficient equal to $2.4$. If the graphene mobility is extremely high, the sample will reach the ballistic regime ($k_B T \tau_0/\hbar \gg1$) and the temperature dependence of $\tau_{\phi}^{-1}$ will turn parabolic \cite{Narozhny02}. Under this circumstance, Eq.~\eqref{eq:MC_SOC} will also be rendered invalid since it is developed for the diffusive regime. However, since our device has a moderate mobility of $\sim 4,000\, \rm{cm}^2\rm{s}^{-1}\rm{V}^{-1}$, it is well in the diffusive transport regime  ($k_B T \tau_0/\hbar \ll 1$); therefore Eq.~\eqref{eq:MC_SOC} is applicable. At low temperatures, the dephasing time $\tau_{\phi}^{-1}$ appears to start deviating from the straight line. In principle, at low temperatures the electron-electron scattering may not be the dominant inelastic scattering mechanism, as compared with electron-phonon interactions, the spin-flip scattering of electrons from localized spins \cite{McCann12}, etc.

%%%%%%%%%%%%%%%%%%%%%%%%%%%%%%%%%%%%%%%%%%%%%%%%%%%%%%%%%%%%%%%%%%%%%%%%
\section{Universal conduction fluctuations (UCF)}

\begin{figure}[H]
\centering
\includegraphics[width=\linewidth]{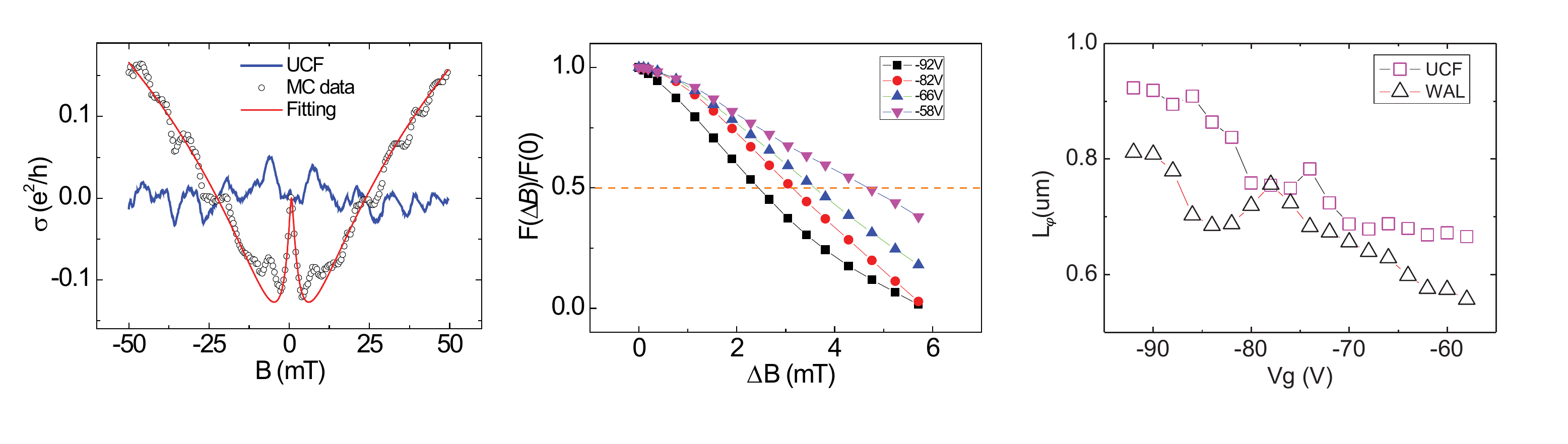}
\caption{(a) UCF extracted from MC by removing the WAL background. (b) Normalized autocorrelation functions at different gate voltages. $F(B_{\phi} )/F(0)=1/2$ gives the characteristic magnetic field $B_{\phi}$. (c) Phase coherence length extracted from UCF and WAL. }
\label{fig:sup_UCF}
\end{figure}

Universal conductance fluctuations can be extracted by removing the WAL background in the MC, as shown in Fig.~\ref{fig:sup_UCF}(a). The WAL curve is fitted by Eq.~\ref{eq:MC_SOC} and describes the experimental data quite well up to 50 mT. In addition to the reproducibility of the MC curves, the nearly symmetric fluctuations in conductance as a function of magnetic field provides further evidence that for UCF.

To calculate the phase coherence length $l_{\phi}$ from the UCF, we utilized the autocorrelation function $F(\Delta B)= \langle \delta \sigma (B+\Delta B)\delta \sigma(B) \rangle$ to find the characteristic magnetic field $B_{\phi}$  $(B_{\phi} l_{\phi}^2=h/2e)$, which is determined by $F(B_{\phi} )=  \frac{1}{2} F(0)$. Figure~\ref{fig:sup_UCF}(b) shows the normalized autocorrelation function at different gate voltages. When the device approaches a higher carrier density, the characteristic field $B_{\phi}$ clearly decreases, indicating an increase in the phase coherence length $l_{\phi}$ due to the weaker electron-electron interaction than the Dirac region. The $l'_{\phi}$ s extracted from UCF agree reasonably well with those extracted from WAL, as shown in Fig.~\ref{fig:sup_UCF}(c), suggesting the validity of Eq.~\eqref{eq:MC_SOC} and thus the extracted spin-orbit scattering rates. 

%%%%%%%%%%%%%%%%%%%%%%%%%%%%%%%%%%%%%%%%%%%%%%%%%%%%%%%%%%%%%%%%%%%%%%%%
\section{DFT Computational details}

Our first-principles calculations were carried out using the projected augmented plane-wave method \cite{Kresse99, Blochl94} as implemented in the Vienna ab initio simulation package. The valence electron configuration for C, W, and S is $2s^2 2 2p^2$, $5d^4 6s^2$, and $3s^23p^4$, respectively. Generalized gradient approximation of Perdew-Burke-Ernzerhof type was used for the description of exchange-correlation interactions among electrons \cite{Perdew96}. Spin-orbit coupling was included in the self-consistent calculation level. We employed $5\times5\times1$ $K$-point grid for Graphene/WS$_2$ heterostructure containing $162$ C, $49$ W and $98$ S atoms. A large supercell was adopted to minimize the lattice mismatch; most of our simulations correspond to a supercell with $9\times 9$ graphene on $7\times 7$ WS$_2$ (see, however, the next section). Slab structures were separated by $\sim 13$ \AA~of vacuum along the surface normal. The energy cutoff for the plane-wave-basis expansion was set to $400 \rm{eV}$. Positions of all atoms were fully relaxed until the convergence of total energies became better than $0.1 \rm{meV}$. In order to treat the van der Waals force properly, we adopted the DFT-D3 method suggested by Ref.~\onlinecite{Grimme10}.  

%%%%%%%%%%%%%%%%%%%%%%%%%%%%%%%%%%%%%%%%%%%%%%%%%%%%%%%%%%%%%%%%%%%%%%%%
\section{Effect of the supercell size on the low-energy states}
\label{app.size_effect}

\begin{figure}[h]
\includegraphics[width=0.5\columnwidth]{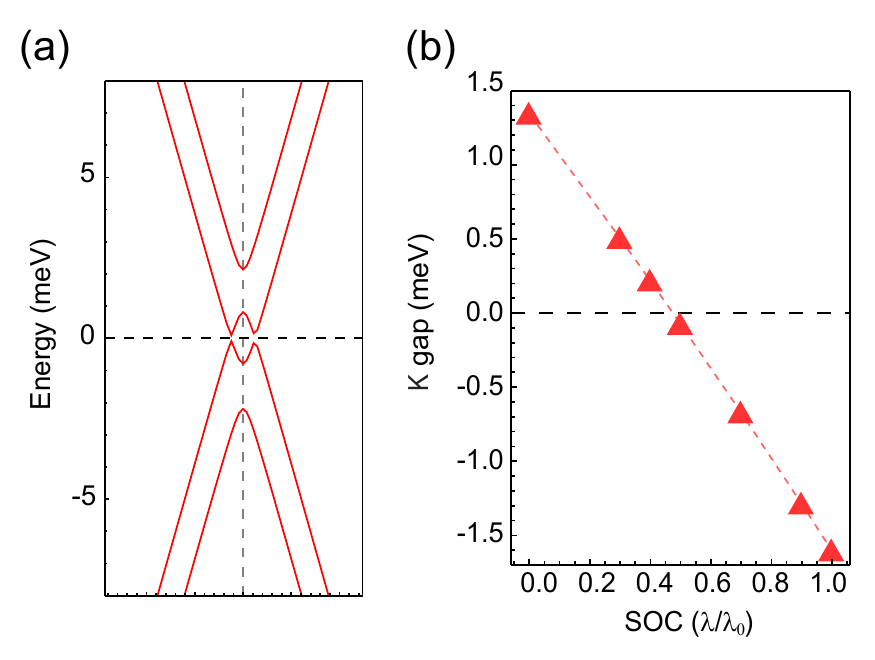}
\caption{(a) Calculated band structure for the slab structure of WS$_2$ monolayer ($4\times4$) on graphene ($5\times 5$). (b) Eenergy gap of the Dirac cone as a function of the spin-orbit coupling strength, which ranges from zero to the true value corresponding to 1 on the horizontal axis. }
\label{fig:sup_band_structure}
\end{figure}

With a $9\times 9$ graphene unit cell, the Dirac cones are folded to $\Gamma$ point.  Consequently, spurious interaction between the valleys may appear in our DFT calculations. To verify that the induced SOC of graphene is not sensitive to the chosen cell size, we repeated the simulations for a different supercell.  Specifically, we adopted $5\times5$ graphene in contact with a $4 \times 4$ WS$_2$ monolayer so that the $K$ and $K'$ valleys reside at distinct momenta.  The lattice mismatch increases to $2.49 \%$ whereas the separation between graphene and WS$_2$ decreases to $3.21$ \AA.  Although we used a different supercell, band energies near the Dirac point remain essentially the same as shown in Fig.~\ref{fig:sup_band_structure}(a). We also found that the Dirac cones are very sensitive to the SOC strength. Figure~\ref{fig:sup_band_structure}(b) shows that the band order becomes inverted at the $K$ and $K'$ points as SOC is turned on from zero to its physical value (we checked that the same band inversion occurs in the $9\times 9$ graphene supercell).  We can therefore conclude that the enhancement of SOC is only weakly affected by the specific atomic configuration and that inter-valley coupling provides a rather minor effect even when the Dirac cones are shifted to the same momentum.  

%%%%%%%%%%%%%%%%%%%%%%%%%%%%%%%%%%%%%%%%%%%%%%%%%%%%%%%%%%%%%%%%%%%%%%%%
\section{Fitting the model Hamiltonian to first-principles calculations}

\begin{figure}[h]
\includegraphics[width=0.5\columnwidth]{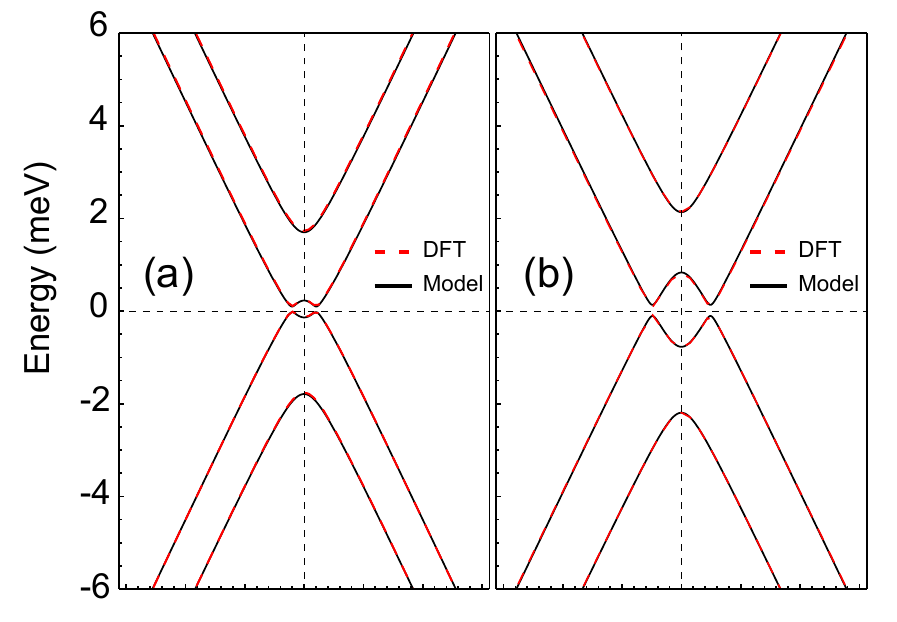}
\caption{Calculated band structure from DFT and the model Hamiltonian for (a) $9\times9$ and (b) $5\times 5$ supercell.}
\label{fig:sub_DFT_fitting}
\end{figure}

By matching energies from the model Hamiltonian in Eq.~\eqref{eq:hamiltonian} with DFT calculations, we determined the following parameters for a  $9\times 9$ ($5\times 5$) graphene supercell: $M=0.79$ $(0.69) \rm{meV}$, $\lambda_{\rm{R}}=0.03$ $(0.08) \rm{meV}$, $\lambda_{\rm{VZ}}=0.96$ $(1.47) \rm{meV}$ and $\lambda_{\rm{I}}\sim 0$ $(0)\rm{meV}$ for the $9\times 9$ ($5\times 5$) supercell. From Fig.~\ref{fig:sub_DFT_fitting} we see that high-quality fits are possible for either interlayer separation.  

\end{document}